\documentclass[aps,pre,twocolumn]{revtex4-1}
\usepackage{amssymb}
\usepackage{amsmath,bm}
\usepackage{graphicx,color}
\usepackage{bbm}
\usepackage{multirow}
\newcommand{\B}[1]{{\bm{#1}}}%% Bold Roman & Greek Lower & Upper Case

\usepackage{hyperref}

\begin{document}
\title{Giant Amplification of Small Perturbations in Frictional Amorphous Solids}
\author{Harish Charan$^1$, Oleg Gendelman$^2$, Itamar Procaccia$^{1,3}$ and Yarden Sheffer$^1$}
\affiliation{$^1$Department of Chemical Physics, The Weizmann Institute of Science, Rehovot 76100, Israel.
$^2$ Faculty of Mechanical Engineering, Technion, Haifa 32000, Israel, 
	 $^3$  Center for OPTical IMagery Analysis and Learning, Northwestern Polytechnical University, Xi'an, 710072 China. }

\begin{abstract}
Catastrophic events in Nature can be often triggered by small perturbations, with ``remote triggering"
of earthquakes being an important example. Here we present a mechanism for the giant amplification of small
perturbations that is expected to be generic in systems whose dynamics is not derivable from a Hamiltonian.
We offer a general discussion of the typical instabilities involved (being oscillatory with an
exponential increase of noise) and examine in detail the normal forms that determine the relevant dynamics.
The high sensitivity to external perturbations is explained for systems with and without dissipation.
Numerical examples are provided using the dynamics of frictional granular matter. Finally we
point out the relationship of the presently discussed phenomenon to the highly topical issue of ``exceptional
points" in quantum models with non-Hermitian Hamiltonians. 
\end{abstract}

\maketitle

\section{Introduction}
\label{intro}

There is growing evidence that remote earthquakes can trigger subsequent large earthquakes with epicenters far
away from the original one, occasionally even far around the world \cite{18OMGB}. While it is obvious that seismic waves
propagate in the crest, intense ones are typically highly damped, and only long wavelength perturbations, which
are relatively weak, can reach long distances. It is then natural to ask, what might be the mechanism for
the amplification of weak perturbations that could be behind this so called ``remote triggering" \cite{13ESKA}.

The aim of this paper is to introduce and discuss a generic mechanism for the giant amplification of small
perturbations in systems whose dynamics is not derivable from a Hamiltonian. The mechanism was discovered
recently in the context of frictional compressed granular matter \cite{19CGPP,19CGPPa,19CCPP,19BCGPZ}, but its relevance appears
more general as is discussed below. We therefore introduce the issue in Sect.~\ref{general} in a very
general setting, touching on fundamental notions of dynamics that are not derivable from a Hamiltonian. We
distinguish between systems in which the forces depend on velocities and those in which they do not. In the
former, the generic oscillatory instability is the Poincare-Andronov-Hopf (PAH) bifurcation, which involves
two modes whose eigenvalues cross the imaginary axis \cite{12MM}. In the latter, the generic mechanism for oscillatory instability
involves {\em four} modes with two pairs of complex conjugate eigenvalues. This second instability is the one
that interests us here, and we show that before its onset the system is highly sensitive to generic external perturbation.
Importantly, the presence of dissipation can turn the interesting instability into a standard PAH bifurcation, and
it is therefore important to assess the role of dissipation, as is done in Subsec.~\ref{effects}. 
In Sect.~\ref{sensitivity} we present the normal forms that allow us to compute how the amplification of 
small noise depends on the frequency of the external noise and on the distance from the instability. We show
that without damping the effects of external noise is giant, diverging as we approach the instability. 
With small damping the amplification is still there, and we compute the maximal amplification as 
a function of the damping strength. In Sect.~\ref{example} we turn to numerical examples. We demonstrate
the amplification mechanism in systems of frictional disks, with and without dissipation. The predictions
of the normal form calculations are tested in details and are vindicated. The last Section offers a summary, 
conclusions, and some comments on the road ahead. 

%%%%%%%%%%%%%%%%%%%%%%%%%%%%%%%%%%%%%%%%%%%%%%%%%%%%%%%%%%%%%%%%%%%%%%%%%%%%%%%%%%%%%%%%%
\section{Instabilities in systems without a Hamiltonian}
\label{general}
\subsection{General setting}
\label{setting}

Consider the very general setting of a system whose degrees of freedom $\bar {\B q}$ obey Newton's equations of motion, but in which the forces are not derivable from
a Hamiltonian. These degrees of freedom can be positions of centers of mass of granules, but also angular degrees of freedom or whatever is necessary to
characterize the state of a given system.  Generically we expect in such cases that it would be possible to separate the forces derivable from a given potential energy, from those forces
that are not. There can be more than one reason why the equations of motion are not derivable from a Hamiltonian. One common reason is the existence
of forces that depend on the velocities $\dot {\bar {\B q}}$ of the degrees of freedom, with the very common example of dissipative terms $-\gamma \dot {\bar {\B q}}$.
Other reasons abound, as will be exemplified below. So quite generally we will consider the equation of motion for the $n$th degree of freedom
\begin{equation} \label{eq1}
m_{n} \ddot{q}_{n} +\frac{\partial V(\bar{q})}{\partial q_{n} } =F_{n}^{np} (\bar{q},\dot{\bar{q}})
\end{equation}
Here $V(\bar{q})$ is the potential energy, $m_{n} $ are the elements of the mass matrix (including if necessary moments of inertia etc.), and
$\bar{F}^{np} (\bar{q},\dot{\bar{q}})$ is the vector of non-potential forces. We assume that the system possesses
a state of equilibrium $\bar{q}=\bar{Q}=const$, that should satisfy:
\begin{equation} \label{eq2}
\left. \frac{\partial V(\bar{q})}{\partial q_{n} } \right|_{\bar{q}=\bar{Q}} =F_{n}^{np} (\bar{Q},0) \ .
\end{equation}
The stability of this equilibrium point can be explored in the state space of the system. Rewriting Eq.~(\ref{eq1}) in terms of the state space, we have
\begin{eqnarray} \label{eq3}
{\bf M}\dot{\bar{S}}&=&\left(\begin{array}{l} {\bar{p}} \\ {-\frac{\partial V(\bar{q})}{\partial \bar{q}} +\bar{f}^{np} (\bar{q},\bar{p})} \end{array}\right) \ , \\~ \nonumber\\
{\bf M}&=&\left(\begin{array}{cc} {\begin{array}{ccc} {m_{1} } & {0} & {0} \\ {0} & {\ddots } & {0} \\ {0} & {0} & {m_{N} } \end{array}} & {0} \\ {0} & {\begin{array}{ccc} {1} & {0} & {0} \\ {0} & {\ddots } & {0} \\ {0} & {0} & {1} \end{array}} \end{array}\right) \ . \nonumber
\end{eqnarray}
We use the following notations
\begin{eqnarray} \label{eq4}
\bar{S}&=&\left(\begin{array}{l} {\bar{q}} \\ {\bar{p}} \end{array}\right);{\rm \; }\bar{f}^{np} (\bar{q},\bar{p})=\bar{F}^{np} (\bar{q},{\bf m^{-1}} \bar{p})\, \\{\bf m^{-1}} &=&\left(\begin{array}{ccc} {m_{1}^{-1} } & {} & {0} \\ {} & {\ddots } & {} \\ {0} & {} & {m_{N}^{-1} } \end{array}\right) \ . \nonumber
\end{eqnarray}
The equilibrium value of the state vector is $\bar{S}_{0} =\left(\begin{array}{l} {\bar{Q}} \\ {0} \end{array}\right)$. Perturbing it as $\bar{S}=\bar{S}_{0} +\bar{\delta }$, one obtains:
\begin{eqnarray} \label{eq5}
{\bf M}\dot{\bar{\delta }}&=&{\bf D}\bar{\delta }+O\left(\left|\bar{\delta }\right|^{2} \right);{\rm \; \; }{\bf D}=\left(\begin{array}{cc} {0} & {I_{N\times N} } \\ {-{\bf H}+\frac{\partial \bar{f}^{np} }{\partial \bar{q}} } & {\frac{\partial \bar{f}^{np} }{\partial \bar{p}} } \end{array}\right)\ , \nonumber \\ {\bf H}&=&\left\| \frac{\partial ^{2} V}{\partial q_{n} \partial q_{m} } \right\| .
\end{eqnarray}
The stability of the state of equilibrium is determined by the solutions of the following eigenvalue problem:
\begin{equation} \label{eq6}
\det \left({\bf D}-{\bf M}\lambda \right)=0  \ .
\end{equation}
The state of equilibrium is stable as long as all the eigenvalues $\lambda$ have negative real parts, $\Re{\lambda}<0$. 
The only apparent generic property of matrix ${\bf D}$ is that it has real entries. Therefore, the loss of stability occurs by two generic scenarios:
\begin{enumerate}
	\item  One of the eigenvalues passes through zero
	\item  A complex conjugate pair passes through the imaginary axis; this is the common Poincare -Andronov - Hopf (PAH) bifurcation.
\end{enumerate}

As said above, these scenarios appear when some forces are not derivable from a Hamiltonian. To proceed, we
discuss the two cases separately. In the first the forces do not depend explicitly on the velocities  $\dot {\bar {\B q}}$;
then the effect of adding the dependence on the velocities will be explored.

\subsection{Non-potential forces that do not depend on velocities}
\label{nonv}

When the non-potential forces are independent of the velocities, i.e. the non-potential forces depend only on coordinates, the stability analysis can be performed in configuration space. Rewriting Equation (\ref{eq1}) in a form
\begin{equation} \label{eq7}
{\bf m}\ddot{\bar{q}}+\frac{\partial V(\bar{q})}{\partial \bar{q}} =\bar{f}^{np} (\bar{q})  \ ,
\end{equation}
we introduce the perturbation $\bar{q}=\bar{Q}+\bar{\Delta }$ and obtain the following linearized problem:
\begin{equation} \label{eq8}
{\bf m}\ddot{\bar{\Delta }}={\bf J}\bar{\Delta }+O\left(\left|\bar{\Delta }\right|^{2} \right);{\rm \; }{\bf J}=-{\bf H}+\frac{\partial \bar{f}^{np} }{\partial \bar{q}}  \ .
\end{equation}
Setting $\bar{\Delta }=\bar{\Delta }_{0} \exp (i\omega t)$, one finally arrives to the following eigenvalue problem:
\begin{equation} \label{eq9}
\det \left({\bf J}-{\bf m}\lambda \right)=0,{\rm \; }\lambda {\rm =}-\omega ^{2} \ .
\end{equation}
The matrix ${\bf J}$ is real, but generically not necessarily symmetric. Without the non-potential forces this matrix becomes the classical Hessian matrix
which is necessarily symmetric. The equilibrium is stable if all eigenvalues $\lambda$ are real and negative. Then, one identifies two possible generic bifurcation scenarios for the loss of stability in this case:
\begin{enumerate}
	\item  A single real eigenvalue passes through zero;
	\item  A pair of negative eigenvalues collide, with the formation of a complex conjugate pair.
\end{enumerate}
It should be stressed that the second scenario substantially differs from the PAH bifurcation, since it requires a minimum of four-dimensional state space.  We will show below that
this bifurcation has huge implications for the our central issue of the giant noise amplification. The reader should note that these results are generic, i.e, they do not depend on the nature of particular models under consideration. In any such case we will have bifurcations with nonzero frequency at the bifurcation point, that will provide the required sensitivity to small perturbations. On the other hand, the need for four simultaneous modes makes this bifurcation
``less generic" than the PAH bifurcation which requires the involvement of only two modes. We explain why this more sensitive bifurcation is nevertheless highly relevant
for noise sensitivity in the next subsection.

\subsection{The effects of velocity dependence}
\label{effects}

The assumption of complete velocity independence is too restrictive for many realistic macroscopic systems,  since some viscous damping is commonly present. Upon first sight, the addition of any amount of viscous damping brings the Eq.~(\ref{eq8}) back to the generic setting Eq.~(\ref{eq1}), with the ``more generic" set of bifurcations. However, if the damping is relatively small and it is possible to consider it as a perturbation, one can still have important consequences of the collision of the pair of negative eigenvalues in the unperturbed system Eq.~(\ref{eq8}).

To illustrate this point, we consider the following generic linear part of the normal form for two degrees of freedom, first without damping:
\begin{eqnarray}
\ddot{q}_1 &+&a_{11} q_1 + a_{12} q_2 =0 \nonumber\\
\ddot q_2 &+& a_{21} q_1 +a_{22} q_2 =0 \ .
\end{eqnarray}
Assuming that $a_{11}$ and $a_{22}$ are both positive, we can rescale time to choose $a_{11}=1$ and then $a_{22}$ will be denoted $\Omega^2$. Then the scale of $q_2$ can be modified
to adjust $a_{12}=1$. Finally $a_{21}$ will be denoted as shown below:
\begin{eqnarray}
\ddot{q}_1 &+& q_1 +  q_2 =0 \nonumber\\
\ddot q_2 &-&\left( \varepsilon +\frac{1}{4}{{(1-{{\Omega }^{2}})}^{2}} \right) q_1 +\Omega^2 q_2 =0 \ .
\end{eqnarray}
Adding now the dissipative terms we obtain finally
\begin{eqnarray}
{{\ddot{q}}}_{1}&+&{{q}_{1}}+{{q}_{2}}+{{\gamma }_{11}}{{{\dot{q}}}_{1}}+{{\gamma }_{12}}{{{\dot{q}}}_{2}}=0 \ , \\
{{{\ddot{q}}}_{2}}&+&{{\Omega }^{2}}{{q}_{2}}-\left( \varepsilon +\frac{1}{4}{{(1-{{\Omega }^{2}})}^{2}} \right){{q}_{1}}+{{\gamma }_{21}}{{{\dot{q}}}_{1}}+{{\gamma }_{22}}{{{\dot{q}}}_{2}}=0
\ . \nonumber
\end{eqnarray}

\begin{figure}
\includegraphics[scale=0.18]{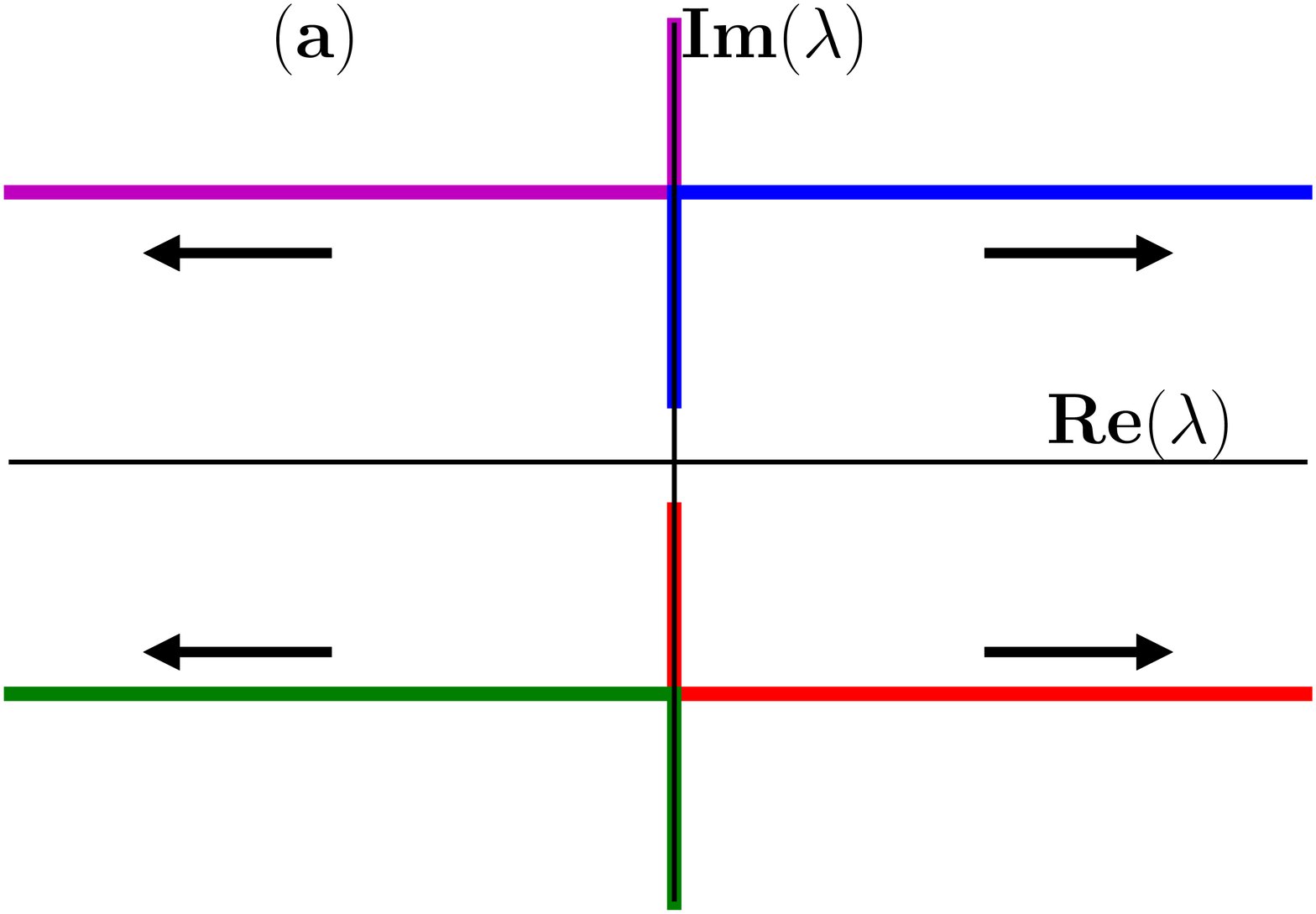}
\includegraphics[scale=0.18]{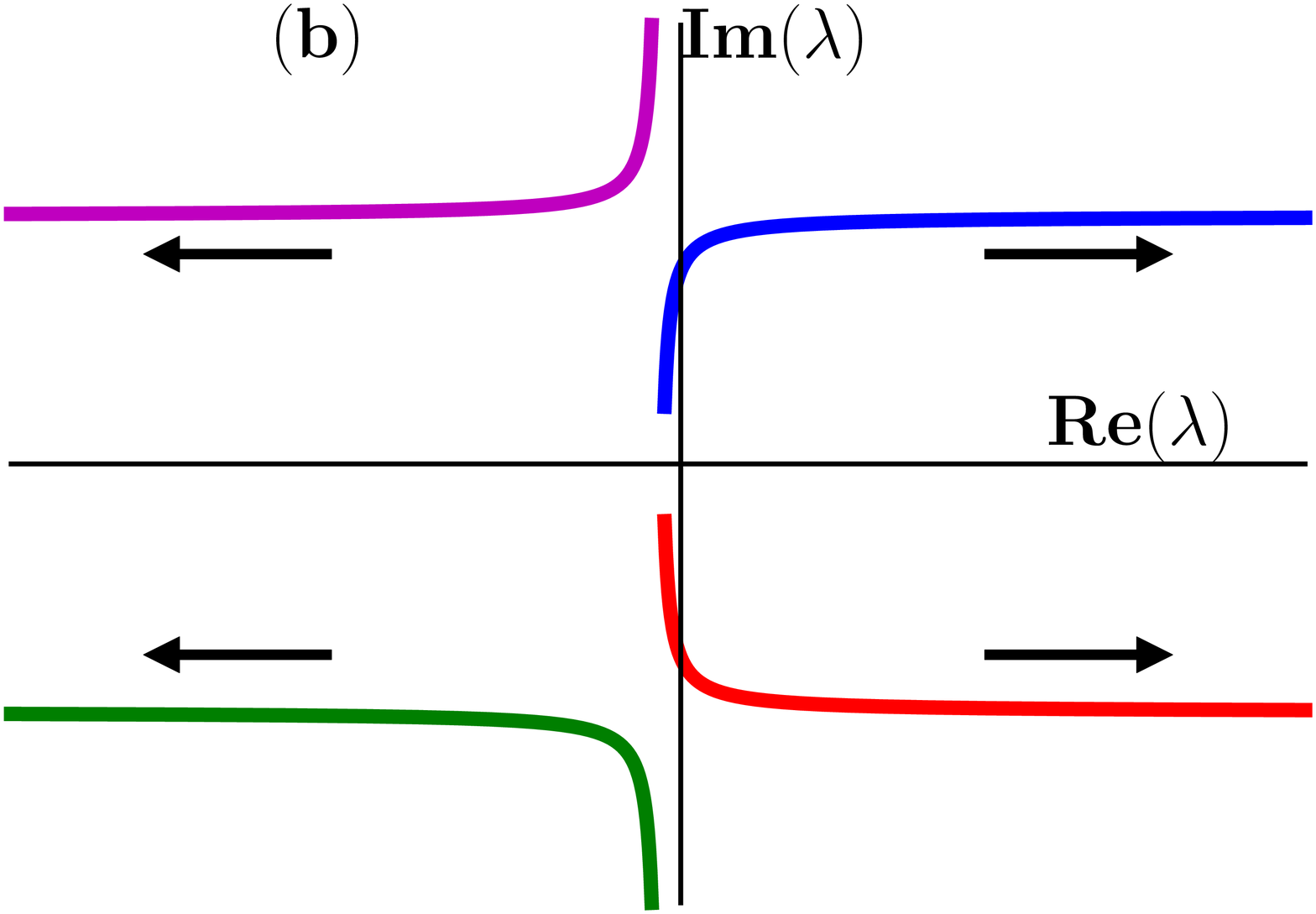}
\caption{Panel a: Evolution of eigenvalues in the case of zero damping. The arrows denote the growth of $\varepsilon$. Panel b: Evolution of eigenvalues in the case of small, but nonzero damping. The arrows denote the growth of $\varepsilon$.}
\label{difference}
\end{figure}

As before, we look for a solution in the form ${q}_{j}={{q}_{j0}}\exp (\lambda t),\text{ }j=1,2$. It is easy to check that in the case of zero damping ${{\gamma }_{kl}}=0$, for $\varepsilon =0$
one obtains a collision of the eigenvalues ${{\lambda }_{1,2}}=-{{\lambda }_{3,4}}=i\sqrt{\frac{1+{{\Omega }^{2}}}{2}}$. The evolution of the eigenvalues with the growth of $ \varepsilon $  in this case is schematically presented in Fig.~\ref{difference} panel a. This is the ideal case of the clean bifurcation involving four modes.
%%%%%%%%%%%%%%%%%%%%%%%%%%%%%%%%%%%%%%%%%%%%%%%%%%%%%%%%%%
%%%%%%%%%%%%%%%%%%%%%%%%%%%%%%%%%%%%%%%%%%%%%%%%%%%%%%%
The inclusion of a small damping changes the flow of the eigenvalues, and they do not collide any more. The flow of the eigenvalues with $\varepsilon$ 
when damping is included is depicted in Fig.~\ref{difference} panel b.
Strictly speaking, one observes the classical PAH bifurcation, with only two eigenvalues crossing the imaginary axis, as expected in the ``more generic" system. However, the other pair of eigenvalues that passes nearby has substantial effect on the dynamics, especially on the sensitivity of the system to external perturbations,  as it will be demonstrated below. The normal
form and the sensitivity to small external perturbations has to be studied with the effect of a small damping as is done explicitly below. 

\section{Normal forms and sensitivity to small perturbations}
\label{sensitivity} 

In this section we present solutions of the normal form equations with external perturbations. The main result of this
section is that there are two mechanisms for enhanced sensitivity to external small broad-band noise. The first is the
usual resonance which is obtained when the external noise includes a frequency which is very close to the natural
frequency of the system mode that is going to become unstable. This resonance is not of a particular interest since
it requires an excitation in the direction of the critical eigenvector. The other more interesting mechanism is induced
by perturbations that are {\em orthogonal} to the critical eigenvectors (which become identical at the instability). 
It is then sufficient to be in a plane
formed by the two colliding eigenvectors, leading to much higher genericity. In addition, in the second case the amplification of the
noise diverges at criticality in the dissipation-less limit (with any frequency of perturbation) and it remains anomalously large
also in the presence of small dissipation. We will first study the undamped case and then add the dissipation.   
 
\subsection{Undamped case}

To develop a normal form for studying the sensitivity to external perturbations,
we conclude from Subsec.~\ref{effects} that the linear equations should include
two independent parameters. We first assume that the equations of motion do not depend
on velocities. Then the most general 2x2
matrix with two independent parameters can be cast in the form of a sum of symmetric and skew-symmetric
matrix.  We choose
the axes so that the symmetric part of the Jacobian matrix is diagonal. Choosing appropriate units
of time the general equation of motion can be written as
\begin{equation}
\partial_{t}^{2}\begin{pmatrix}x\\
y
\end{pmatrix}=- \B J\begin{pmatrix}x\\
y
\end{pmatrix} = -
\begin{pmatrix}1-\delta & \eta\\
-\eta & 1+\delta
\end{pmatrix}\begin{pmatrix}x\\
y
\end{pmatrix}\label{eq:hom}
\end{equation}
with $1>\delta$.

Substituting $\left(x,y\right)=\left(X,Y\right)e^{i\omega t}$ with constant $X,Y$
we find that the eigenfrequencies are obtained as $\omega_{i}=\sqrt{\lambda_{i}}$
with $\lambda_{i}$ being the eigenvalues of the Jacobian matrix $\B J$:
\begin{equation}
\label{deflam}
\lambda_{1,2}=1\mp\delta\sqrt{1-\mu^{2}},
\end{equation}
where $\mu=\frac{\eta}{\delta}$. Clearly the system develops a complex
pair of eigenvalues for $\mu>1$.  We now take $\epsilon=1-\mu=1-\frac{\eta}{\delta}$ and, in order
for the system to be critical, assume that $\epsilon\ll1$. Using Eq.~(\ref{deflam}), the associated
frequencies are to leading order
\begin{equation}
\omega_{1,2} \approx 1 \mp \delta \sqrt{\epsilon/2}\ .
\end{equation}

The eigenvectors $\tilde{\B v}_{1,2}=\left(X,Y\right)$
are obtained as
\begin{equation}
\tilde {\B v}_{1,2}=\begin{pmatrix}1\pm\sqrt{1-\mu^{2}}\\
\mu
\end{pmatrix}.\label{eq:eigvec}
\end{equation}
In the limit $\mu\to 1$ the two critical eigenvectors coincide and become (1,1). 
 The two eigenvectors (\ref{eq:eigvec})
then become, after normalizing such that $\B v_{1,2} \equiv \tilde {\B v}_{1,2}/\left|\tilde {\B v}_{1,2}\right|=1+O\left(\sqrt{\epsilon}\right)$,
\begin{equation}
\B v_{1,2}=\frac{1}{\sqrt{2}}\begin{pmatrix}1\pm\sqrt{\epsilon/2}\\
1\mp\sqrt{\epsilon/2}
\end{pmatrix}.\label{eq:eigvecnorm}
\end{equation}

The system has two mechanisms for amplifying outside noise: First, when
the frequency of the outside noise is close to $\omega_{1,2}$ we
obtain a solution whose amplitude is proportional to $\left|\omega-\omega_{1,2}\right|^{-1}$;
this is a simple resonance mechanism and has nothing to do with our criticality.
Second, from Eq.~(\ref{eq:eigvecnorm}) we can observe that near the critical
point the two eigenvectors become parallel. As a result, an initial
condition $(x_0,y_0)$ orthogonal to $\boldsymbol{v^{*}}=\left(1,1\right)/\sqrt{2}$
will result in oscillations whose amplitude diverges as $\epsilon^{-1/2}$.

The traditional resonance is obtained as a particular solution of the following equation in which the external perturbation can have  any arbitrary frequency $\omega$:
\begin{equation}
\partial_{t}^{2}
\begin{pmatrix}x\\
y
\end{pmatrix}=
-\begin{pmatrix}1-\delta & \eta\\
-\eta & 1+\delta
\end{pmatrix}\begin{pmatrix}x\\
y
\end{pmatrix}+\boldsymbol{f}\cos\omega t\label{eq:forcing}
\end{equation}
with $\boldsymbol{f}=F\left(1,-1\right)/\sqrt{2}$ chosen so that
$\boldsymbol{f}\bot\boldsymbol{v^{*}}$.

Substituting $\left(\xi,\zeta\right)=\left(X,Y\right)\cos \omega t$,
we get the particular solution
\begin{align}
\begin{aligned}
\begin{pmatrix}\xi\\
\zeta
\end{pmatrix}& =\frac{F}{\sqrt{2}\left(\omega^{2}-\lambda_{1}\right)\left(\omega^{2}-\lambda_{2}\right)}\\
 & \times\begin{pmatrix}1-\omega^{2}+\delta\left(2-\epsilon\right)\\
-1+\omega^{2}+\delta\left(2-\epsilon\right)
\end{pmatrix}\cos\omega t \ .
\end{aligned}
\label{eq:par}
\end{align}
As expected, with identifying $\omega_i=\pm \sqrt{\lambda_i}$, the amplitude of this particular solution
diverges as
\begin{equation}
A_{\rm par} \sim F\left|\omega-\omega_{1,2}\right|^{-1} \ ,
\label{eq:undamped_response}
\end{equation}
 as a result of the resonance.

The more interesting and important mechanism for noise amplification is associated
with the {\em homogeneous} solution of Eq. (\ref{eq:forcing}).
 This solution is identified by choosing the initial
displacement $\left(x\left(0\right),y\left(0\right)\right)=-(\xi(0),\zeta(0))$,
which annuls the particular solution.

\begin{figure}
\includegraphics[width=0.4\textwidth]{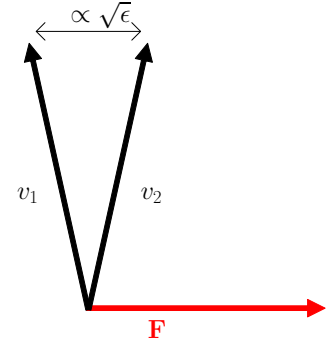}
\caption{The eigenvectors are close to coalescence
when criticality is approached. The external force is perpendicular
to them.}
\label{fig:The-geometry-of}
\end{figure}

The general form of the homogeneous solution is
\begin{equation}
\boldsymbol{x}_{{\rm hom}}=\sum_{i=1,2}a_i\boldsymbol{v}_{i}\cos\omega_{i}t \ ,
\end{equation}
where $a_i$ are determined from the equation
\begin{equation}
\sum_{i}a_i\boldsymbol{v}_{i}=\begin{pmatrix}-\xi(0)\\
-\zeta(0) 
\end{pmatrix} \ .
\end{equation}
We will show here that if the angle between $\boldsymbol{v}_{1}$ and
$\boldsymbol{v}_{2}$ is small ($\boldsymbol{v}_{1}\cdot\boldsymbol{v}_{2}=1+O(\epsilon)$),
an initial condition perpendicular to $\boldsymbol{v}_{i}$ will result
in divergences of $a_i$ as $\left|a_i\right|\propto\epsilon^{-1/2}$
(see illustration in figure \ref{fig:The-geometry-of}). Namely, solving
(\ref{eq:hom}) with $\boldsymbol{x}_{{\rm hom}}\left(0\right)=\left(1,-1\right)/\sqrt{2}$
we obtain
\begin{equation}
\begin{aligned}x_{{\rm hom},\bot} & \approx \begin{pmatrix}\frac{1}{2\sqrt{\epsilon}}+\frac{1}{2\sqrt{2}}\\
\frac{1}{2\sqrt{\epsilon}}-\frac{1}{2\sqrt{2}}
\end{pmatrix}\cos\left(\omega_{1}t\right)\\
 & -\begin{pmatrix}\frac{1}{2\sqrt{\epsilon}}-\frac{1}{2\sqrt{2}}\\
\frac{1}{2\sqrt{\epsilon}}+\frac{1}{2\sqrt{2}}
\end{pmatrix}\cos\left(\omega_{2}t\right)\\
 & \approx\frac{1}{\sqrt{\epsilon}}\begin{pmatrix}1\\
1
\end{pmatrix}\sin\left(t\right)\sin\left(\frac{\delta \epsilon^{1/2}}{\sqrt{2}}t\right)\\
 & +\frac{1}{\sqrt{2}}\begin{pmatrix}1\\
-1
\end{pmatrix}\cos\left(t\right)\cos\left(\frac{\delta \epsilon^{1/2}}{\sqrt{2}}t\right).
\end{aligned}
\label{eq:xhom}
\end{equation}
%%%%%%%%%%%%%%%%%%%%%%%%%%%%%%%%%%%%%%%%%%%%%%
%%%%%%%%%%%%%%%%%%%%%%%%%%%%%%%%%

Two important conclusions are to be drawn here: First, the
amplitude $A$ of the homogeneous solution goes as
\begin{equation}
A\sim\epsilon^{-1/2}.\label{ampeps}
\end{equation}
Second, we find a new time constant $\tau_d$ for the divergent solution to become significant, or the
proper time for the solution to reach the maximal amplitude. From
(\ref{eq:xhom}) we obtain
\begin{equation}
\tau_d \sim \frac{1}{\delta\sqrt{\epsilon}}\sim \frac{1}{|\omega_1-\omega_2|}.
\end{equation}
In particular, the time for the maximal amplitude to be reached diverges
as $\epsilon\rightarrow0$. This leads to the final solution for the
amplitude gain, taking (\ref{eq:par}) and (\ref{eq:xhom}) we obtain
\begin{equation}
A_{{\rm max}}\sim\frac{\epsilon^{-1/2}}{\left|\omega-1\right|}F.\label{eq:scaling_undamped}
\end{equation}
This holds true for $\omega\approx \omega_{1}$ or $\omega\approx\omega_{2}$. Note that for any arbitrary
frequency one still has the divergence proportional to $\epsilon^{-1/2}$. 

The reader should note that when this normal form is embedded in a {\em nonlinear} system,
the increase in oscillations can easily ignite the nonlinear terms and drive the
system further from equilibrium. Thus one may not see the linear blow-up in its entirety
because nonlinearities will become dominant. An example will be shown in Sect.~\ref{example}.

\subsection{The effects of damping}

Let us now consider the case in which a damping force
\begin{equation}
\boldsymbol{f}_{{\rm damp}}=-\frac{1}{\tau}\partial_{t}\begin{pmatrix}x\\
y
\end{pmatrix}
\end{equation}
is added to equation (\ref{eq:forcing}). How is the scaling relation
(\ref{eq:scaling_undamped}) expected to change?

For the particular solution with the standard resonance, the problem reduces to a simple damped-driven oscillator,
which results in a Lorentzian response, so that Eq.~(\ref{eq:undamped_response})
will become
\begin{equation}
A_{{\rm par}}\sim\frac{1}{\sqrt{\omega^{2}/\tau^{2}+\left(\omega^{2}-\omega_{1,2}^{2}\right)^{2}}}F \ .
\end{equation}
Similarly to Eq.~(\ref{eq:undamped_response}) for $\omega-\omega_{1,2}\gg 1/\tau$. 
The response will achieve a maximal value that is evaluated as
\begin{equation}
A_{\rm par,\rm max}\sim {F\tau}\sim FQ \ .
\label{amppar}
\end{equation}
where $Q$ is the quality factor of the system at criticality.

For the homogeneous solution (\ref{eq:xhom}), we note that $\boldsymbol{v}_{1,2}$
from equation (\ref{eq:eigvecnorm}) are still the eigenvectors
of the system near criticality. A homogeneous solution of the form
$\left(X,Y\right)e^{i\omega t}$ will now solve
\begin{equation}
\left[-\begin{pmatrix}\omega^{2}-i\omega/\tau & 0\\
0 & \omega^{2}-i\omega/\tau
\end{pmatrix}+\begin{pmatrix}1-\delta & \eta\\
-\eta & 1+\delta
\end{pmatrix}\right]\begin{pmatrix}X\\
Y
\end{pmatrix}=0 \ ,
\end{equation}
so that $\boldsymbol{v}_{1,2}$ will still be obtained from (\ref{eq:eigvecnorm}).
In the case of a small damping ($1/\tau \ll 1$) Eq.~(\ref{eq:xhom})
will then become
\begin{equation}
\begin{aligned}x_{{\rm hom}} & \approx\frac{1}{\sqrt{\epsilon}}\begin{pmatrix}1\\
1
\end{pmatrix}\sin\left(t\right)\sin\left(\frac{\delta \epsilon^{1/2}}{\sqrt{2}}t\right)e^{-\frac{t}{\tau}}\\
 & +\frac{1}{\sqrt{2}}\begin{pmatrix}1\\
-1
\end{pmatrix}\cos\left(t\right)\cos\left(\frac{\delta \epsilon^{1/2}}{\sqrt{2}}t\right)e^{-\frac{t}{\tau}} \ .
\end{aligned}
\end{equation}
The maximal amplitude is obtained as
\begin{equation}
A\sim\max_{t}\left[\frac{1}{\sqrt{\epsilon}}\sin\left(t/\tau_d\right)e^{-\frac{t}{\tau}}\right] \ ,
\end{equation}
which can be approximated near criticality, i.e. for $\tau_d\gg \tau$, as
\begin{equation}
\begin{aligned}A & \sim\max_{t}\left[\delta te^{-\frac{t}{\tau}}\right]\\
 & \sim\delta\tau=\delta Q \ .
\end{aligned}
\label{amptau}
\end{equation}
This means that the scaling
(\ref{ampeps}) has a point of saturation as $\epsilon\rightarrow0$,
which is proportional to the quality factor. From Eq.~(\ref{amppar})
and Eq.~(\ref{amptau}) we finally obtain for $|\omega-\omega_{1,2}|\ll 1/\tau$ and $\tau_d\gg \tau$,
\begin{equation}
A_{{\rm max}}\sim Q^{2}\delta F \ ,
\label{ampfinal}
\end{equation}
which is our main result.

Note that in the opposite limit, i.e. if $\tau_d\ll \tau$, the damping is not strongly effective
and we expect and recover Eq.~(\ref{ampeps}). 

The conclusion is that while the amplification due to the regular resonance is bounded by the
quality factor, here we have amplification by the square of the quality factor. This is likely
to bring the system into the nonlinear regime where the response can spontaneously grow further, and
sometime catastrophically. 
%%%%%%%%%%%%%%%%%%%%%%%%%%%%%%%%%%%%%
\begin{figure}
	%\vskip 0.4 cm
	\includegraphics[width=0.38\textwidth]{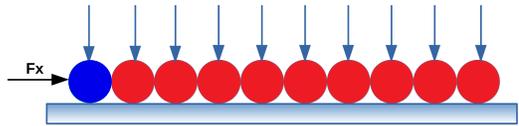}
	\caption{The model consists of $N$ identical disks (here and in the simulations below $N=10$) which interact via
		Hertz and Mindlin forces between themselves and the substrate below. A constant force $F_y$ is applied
		to press them against the substrate, and an external force $F_x$ is applied to the first disk, increasing it
		quasistatically until a pair of complex eigenvalues gets born, signaling an oscillatory instability. From that point on the Newtonian dynamics takes the system from static to dynamical friction. }
	\label{model}
\end{figure}
%%%%%%%%%%%%%%%%%%%%%%%%%%%%%%

\section{Example: frictional disks in 2 dimensions}
\label{example}

As an example we choose a system that was studied in detail to demonstrate a transition from static to dynamical friction \cite{19CCPP}, see Fig~\ref{model}. 
It consists of $N$ 2-dimensional disks of radius R, with their initial center of mass positioned at $x_i=(2i -1)R; y_i=R$, $i=1\cdots N$, aligned over an infinite substrate at $y=0$. Each disk is pressed with an identical force $F_y$ normal to the substrate, providing a very simple model of asperity contacts in more realistic systems. The boundary conditions are periodic such that
the disk $i=N$ is in contact with the disk $i=1$. The disk-disk and and disk-substrate interactions are the time honored Hertz and Midnlin forces that are not derivable
from a Hamiltonian. These forces 
are described in detail in Ref.~\cite{19CCPP}. Forces and torques are annulled by force minimization protocol to reach mechanical equilibrium. After attaining equilibrium we increase quasistatically a force $F_x$ which is applied at the center of mass of the disk $i=1$. At some critical value of $F_x=F_{x,c}$  the system becomes
unstable with respect to an oscillatory instability \cite{19CGPPa,19CGPP}. This instability can trigger a transition from static to dynamical friction. It was
demonstrated before that once the system is in the unstable regime, even numerical noise can trigger the instability, and the response can exhibit signal
increase of twenty orders of magnitude. The aim of this section is to study the sensitivity of the system when it is still in the {\em stable} regime. We will 
demonstrate extreme sensitivity with giant response to small perturbations, triggering the instability also when in the absence of perturbations the system is completely
stable .
%%%%%%%%%%%%%%%%%%%%%%%%%%%%%%%%%%%%%
\begin{figure}
	\includegraphics[width=0.38\textwidth]{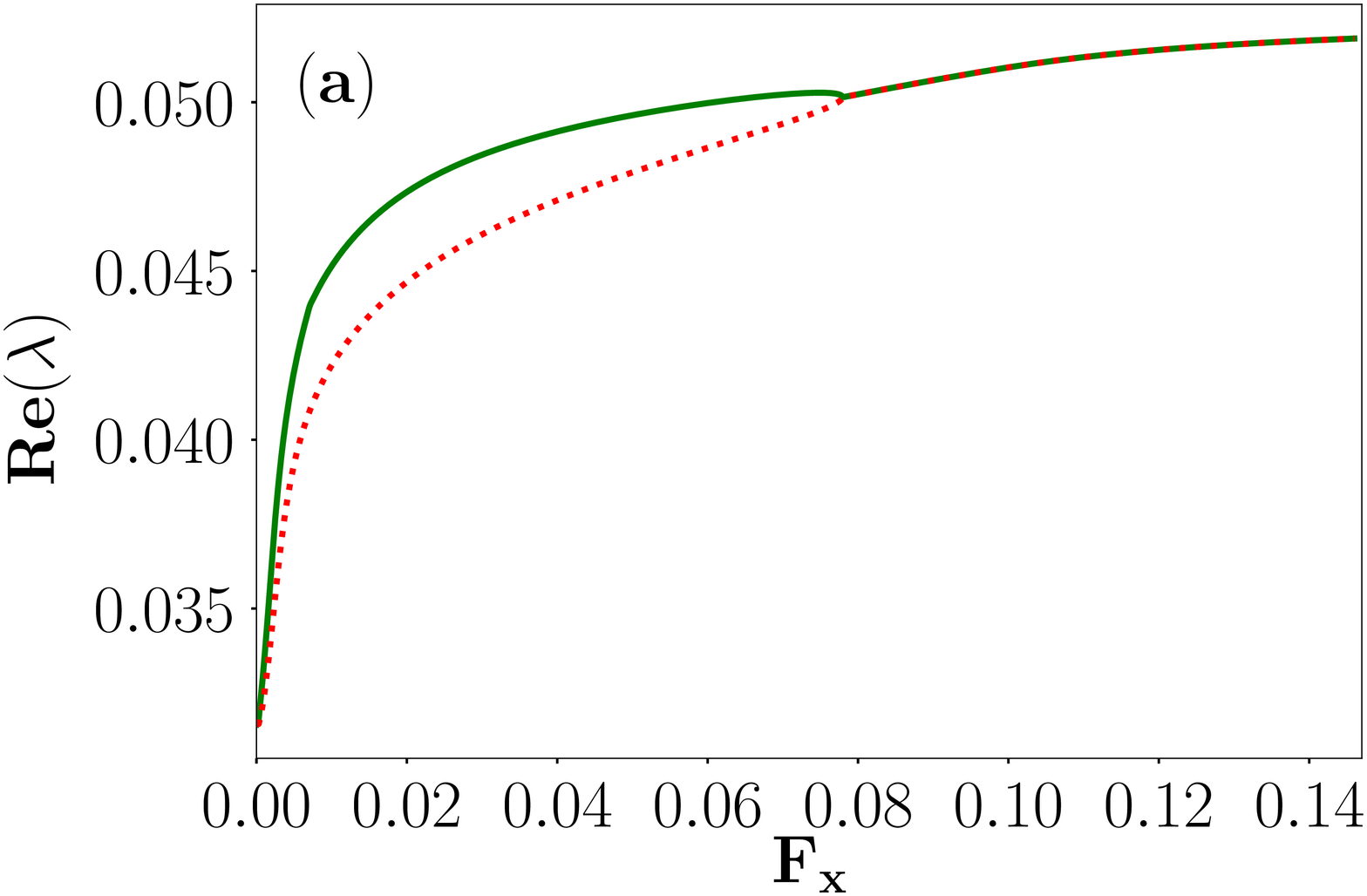}
	\includegraphics[width=0.38\textwidth]{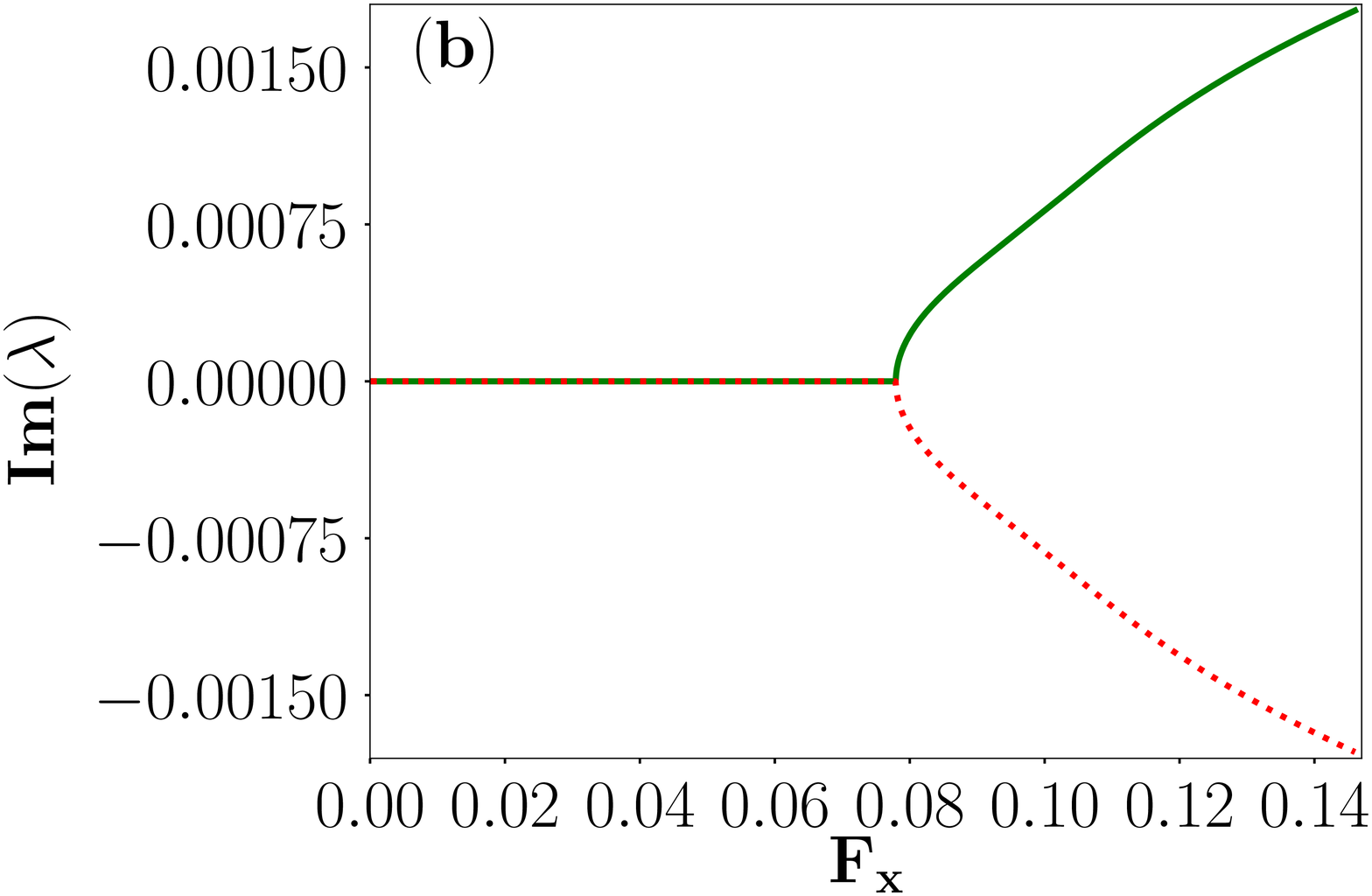}
	\caption{The bifurcation diagram. At $F_x=F_{x,c} \approx 0.0779$ the two real eigenvalues coalesce
		and the two imaginary parts bifurcate.}
	\label{eigenvalues}
\end{figure}
%%%%%%%%%%%%%%%%%%%%%%%%%%%%%%%%%%%%%%%%%%%%%%%%

The dynamics are Newtonian; denoting the set of coordinates $\B q_i=\{\B r_i, \B \theta_i\}$:
\begin{eqnarray}
\label{Newton2a}
m\frac{d^2 \B r_{i}}{dt^2}&=&{\B F}_i(\B q_{i-1},\B q_i,\B q_{i+1})\ , \quad q_{N+1} = q_1 \ ,\\
\label{Newton2b}
I\frac{d^2 \B \theta_{i}}{dt^2}&=&{\B T}_i(\B q_{i-1},\B q_i,\B q_{i+1})\ ,
\end{eqnarray}
where $m$ and $I$ are the mass and moment of inertia of the $i$th disk, ${\B F}_i$ and ${\B T}_i$ are the total force and the torque on disk $i$ respectively. Time is measured in units
of $\sqrt{m2Rk_n}$ and length in units of $2R$. To study the effect of dissipation we add to the RHS of Eq.~(\ref{Newton2a}) a term $-\dot{ \B  r_i}/\tau$. We will study
the effect of decreasing $\tau$ (increasing dissipation) on the sensitivity of the system to small external perturbations. 
%%%%%%%%%%%%%%%%%%%%%%%%%%%%%%%%%%%%%
\begin{figure}
	\includegraphics[width=0.32\textwidth]{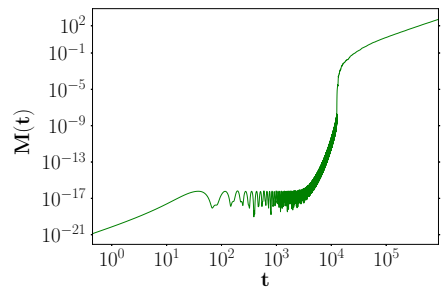}
	\caption{Unstable dynamics after the bifurcation of the imaginary parts of the eigenvalues. Here $F_x=0.086$ and 
		numerical noise is sufficient to bring about a huge growth in the MSD, more than twenty orders of magnitude.}
	\label{unstable}
\end{figure}
%%%%%%%%%%%%%%%%%%%%%%%%%%%%%%%%%%%%%%%%%%%%%%%%%
The critical point $F_{x,c}$ is identified as the point of coalescence of two real eigenvalues and the bifurcation of imaginary part, cf. Fig.~\ref{eigenvalues}.
Without damping, when $F_x$ exceeds the critical value  $F_{x,c}$ the system is always unstable, and it suffices to have numerical noise to develop the
instability. To see the effect on the instability, we measure the mean-square displacement (MSD) defined here as
\begin{equation}
M(t)\equiv \frac{1}{10}\sum_{i=1}^{10} |\B r_i(t)-\B r_i(0)|^2 \ .
\label{MSD}
\end{equation}
The development of this quantity in the unstable regime is exemplified in Fig.~\ref{unstable}. One sees that just with numerical noise the response increases
by more than twenty orders of magnitude and the system exhibits a transition from static to dynamic scaling, where the MSD is growing exponentially in time.

\begin{figure}
	\includegraphics[width=0.28\textwidth]{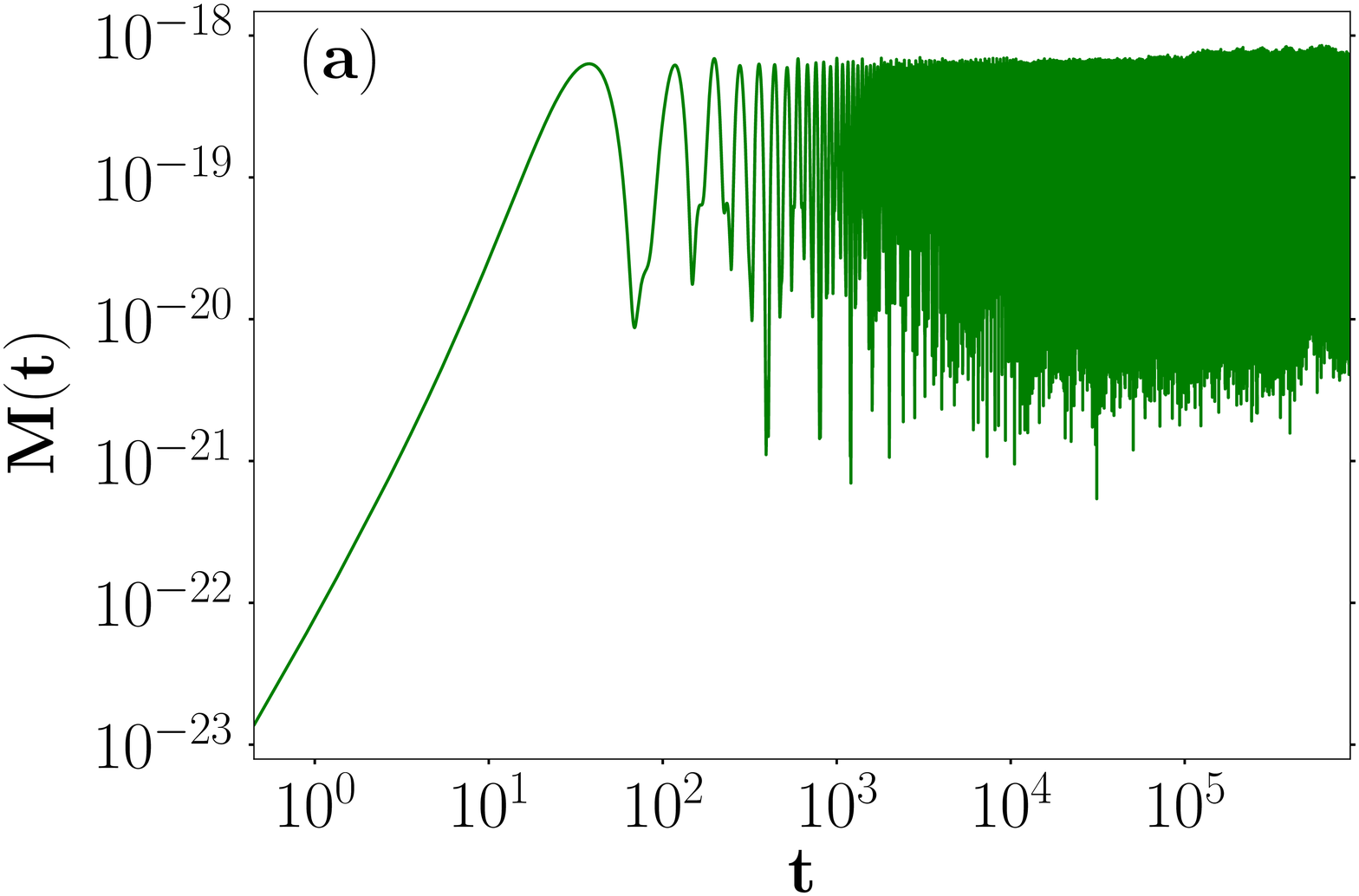}
	\includegraphics[width=0.28\textwidth]{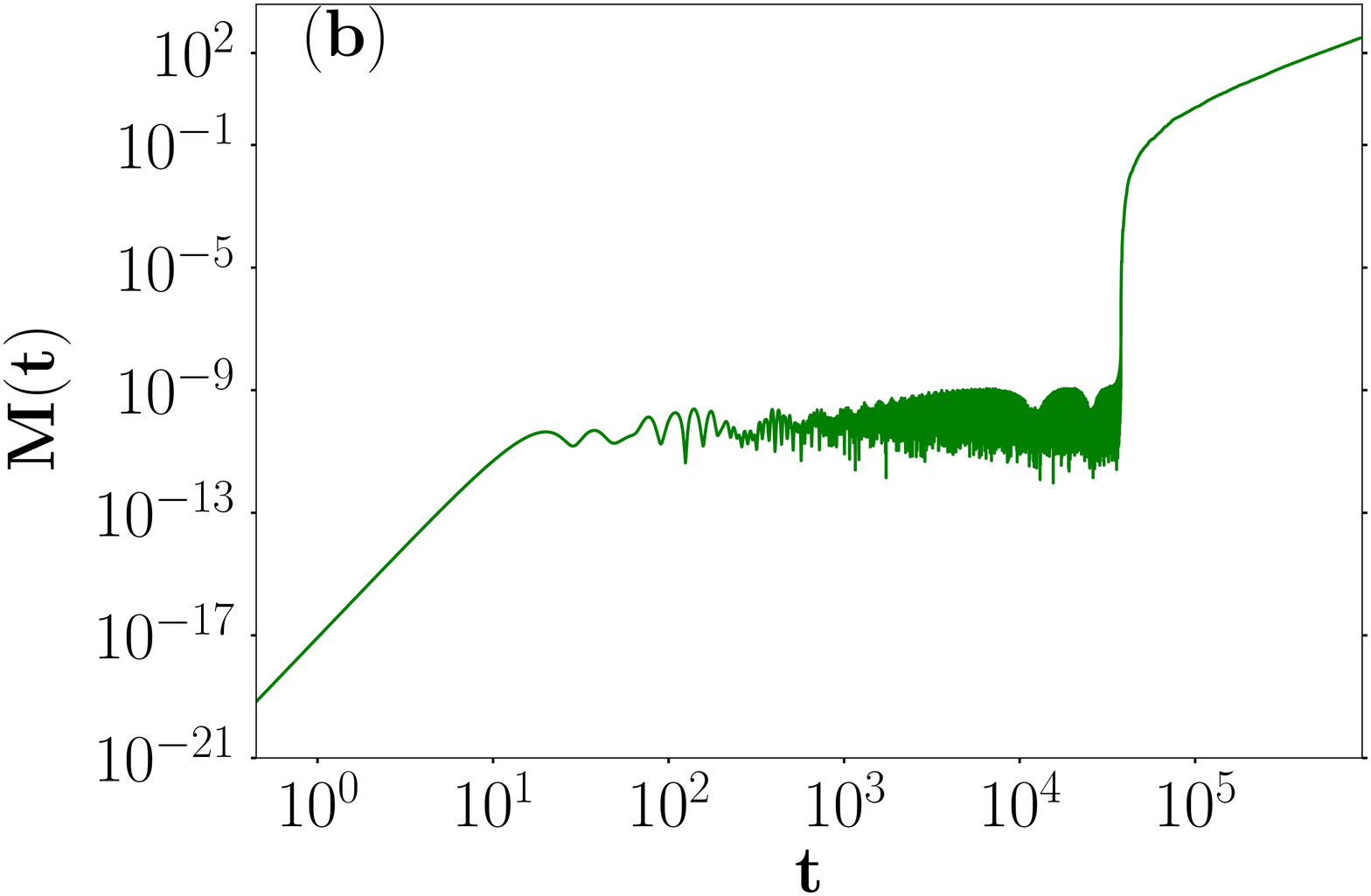}
		\includegraphics[width=0.28\textwidth]{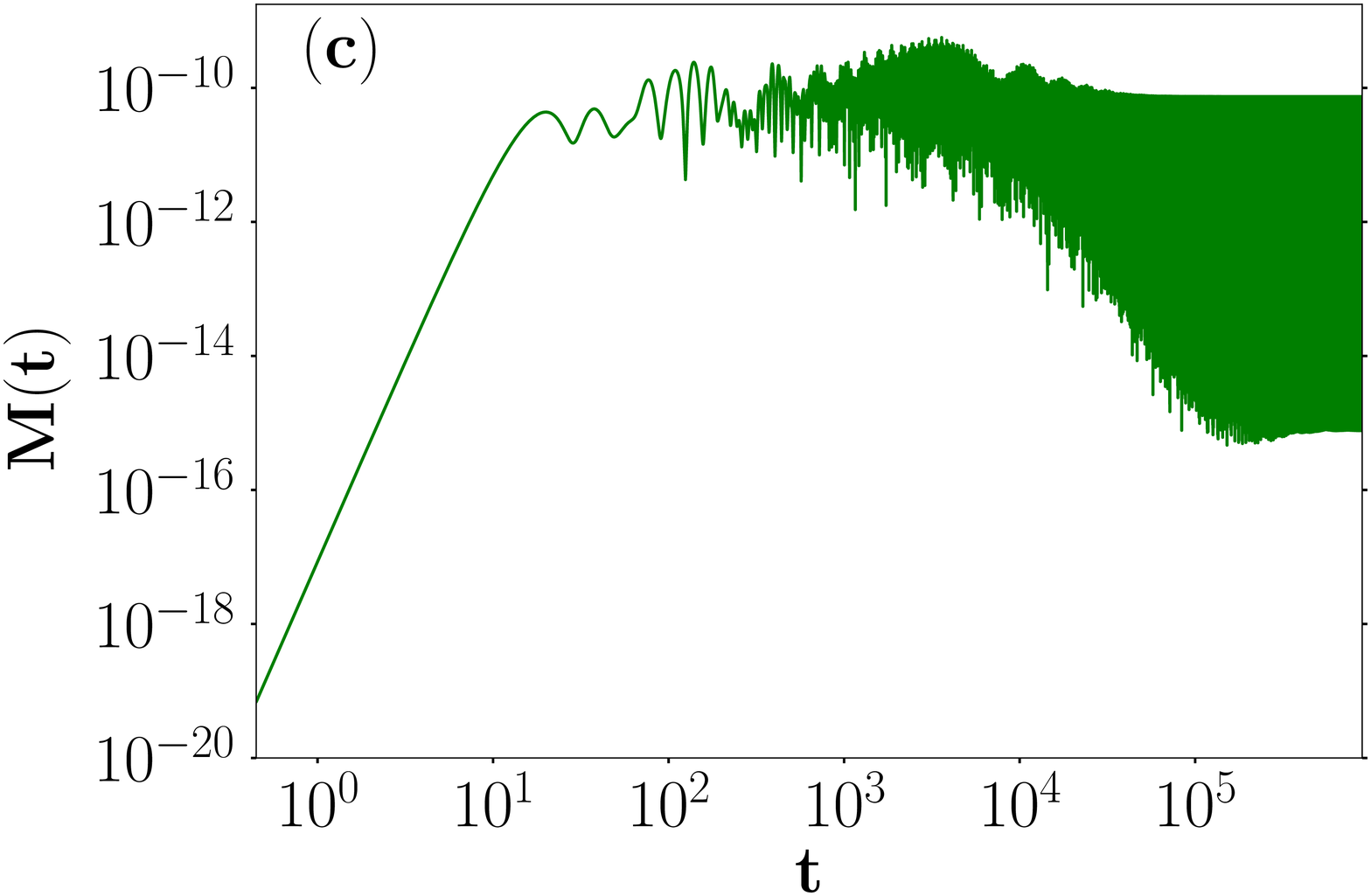}
	\caption{Panel a: with $\epsilon\approx 0.058$ and without perturbation and without damping the systems is stable, never
	reaching any substantial value of MSD. Panel b: for the same value of $\epsilon$ with $\alpha=0.005$ and $\omega=0.1$
and without damping the small perturbation is sufficient to throw the system unstable. Panel c: with $\tau=100$ and the same perturbation the
MSD grows by more than nine orders of magnitude but the instability is not switched on.}
\label{lovely} 
\end{figure}

To demonstrate the sensitivity associated with this instability, we will apply now small perturbations at given forces $F_x<F_{x,c}$ and will monitor the response for different values of $\epsilon$ and $\tau$. The problem is now
multidimensional (3N-dimensional), so we extract $\epsilon$ from the scalar product of the two eigenvectors that coalesce at $F_x=F_{x,c}$. Having
Eq.~(\ref{eq:eigvecnorm}) in mind, and continuing to denote the two coalescing vectors as $\B v_1$ and $\B v_2$ (out of the 3N available eigenvectors), we define 
\begin{equation}
\epsilon \equiv 2(1-\B v_1\cdot \B v_2)\ .
\label{defeps}
\end{equation}
The perturbation will be taken in the form 
\begin{equation}
\B f cos(\omega t) \equiv\alpha(\B v_2-\B v_1) \cos (\omega t) \ ,
\label{forcing}
\end{equation}
with small $\alpha$ and $\omega$ far from resonance. 
In Fig.~\ref{lovely} we then demonstrate how, for $\epsilon\approx 0.058$ the dynamics is stable in the absence of forcing and
damping. This is seen in panel (a). The dynamics remains for ever in the numerical noise level of order $10^{-18}$. 
%%%%%%%%%%%%%%%%%%%%%%%%%%%%%%%%%%%%%
%%%%%%%%%%%%%%%%%%%%%%%%%%%%%%%%%%%%%%%%%%%
In panel (b) we demonstrate the response of the MSD to a small perturbation with $\alpha=0.005$ and $\omega=0.1$,
far away from resonance. Without damping the effect of the small perturbation is so huge that the system
is kicked over to the unstable regime and the transition to dynamic friction takes place. Note that the MSD shoots
up by more than 20 orders of magnitude, making manifest the huge sensitivity to small perturbations. Panel (c) demonstrates
the effect of damping. With $\tau=100$ the transition to dynamic friction is avoided, but the response of
the MSD is eight orders of magnitude larger than that obtained without the small perturbation! 

\begin{figure}
\includegraphics[width=0.32\textwidth]{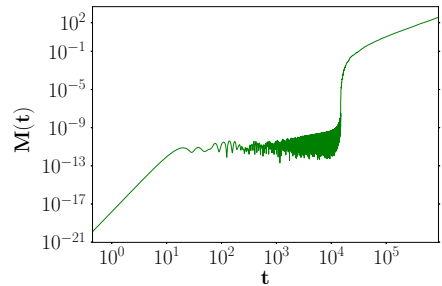}
\caption{The effect of coming closer to the onset of instability. Here the forcing amplitude, $\omega$ 
and $\tau$ are the same as in panel (c) of Fig.~\ref{lovely} with the only difference that 
$\epsilon\approx 0.01$ instead of 0.058. This is enough to increase the response to a level that is 
picked up by the nonlinearities to trigger the instability.}
\label{show}
\end{figure}

Is damping always saturating the instability? Of course not. In Fig.~\ref{show} the results  of repeated simulations
with the strength of forcing as in panel (c) of Fig.~\ref{lovely} with the only difference that now $\epsilon\approx 0.01$
In other words, the system is in the stable regime but closer to the instability. 
%%%%%%%%%%%%%%%%%%%%%%%%%%%%%%%%%%%%%
%%%%%%%%%%%%%%%%%%%%%%%%%%%%%%%%%%%%%%%%%%%%%%%%%%%%%%%%%%%%%%
Now the damping fails to delay the onset of the instability. The growth in the amplitude of oscillations is 
large enough to trigger the onset of instability as the nonlinearities kick in. Note
that we kept the value of $\alpha$ fixed but $|\B v_1-\B v_2|$ is reduced here, so the noise amplitude is smaller
than before. Nevertheless, the instability was triggered in spite of the damping due to the enhanced amplitude of the response. 
%%%%%%%%%%%%%%%%%%%%%%%%%%%%%%%%%%%%%
\begin{figure}
	\includegraphics[width=0.32\textwidth]{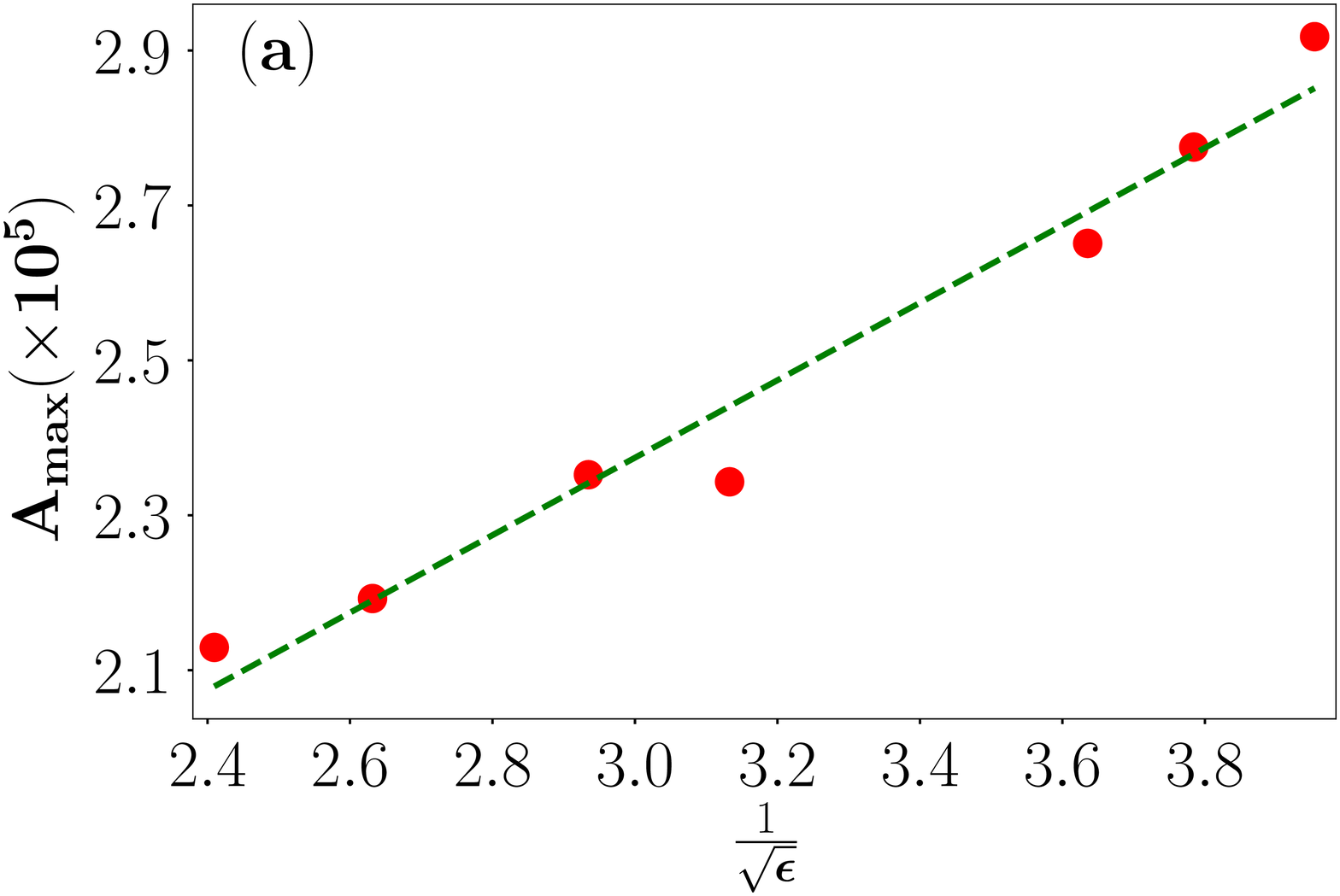}
	\includegraphics[width=0.32\textwidth]{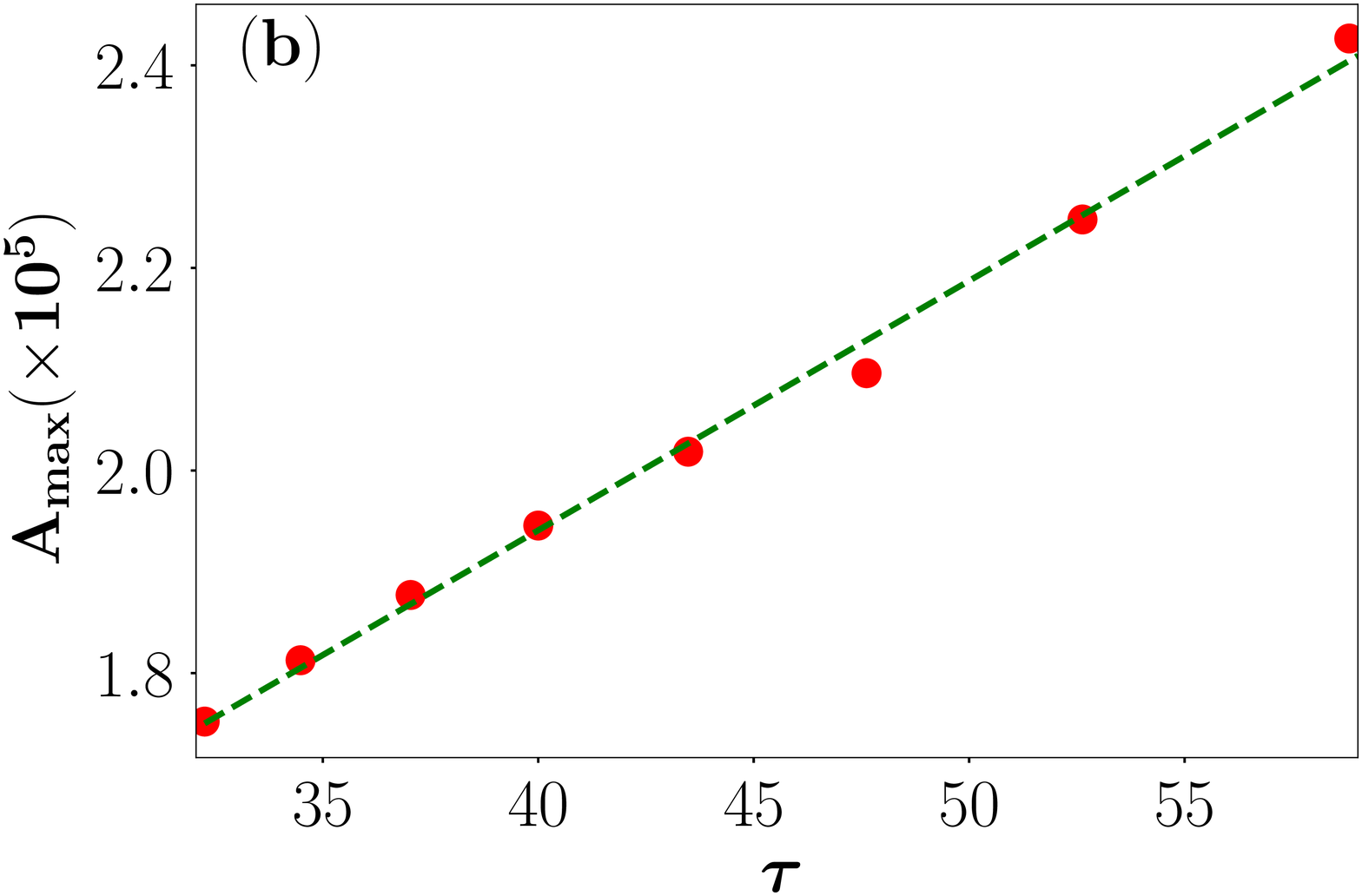}
	\caption{The dependence of the amplitude of the response on the distance from instability and damping. Panel a: $A_{\rm max}$ as a function of $1/\sqrt{\epsilon}$ for fixed $\tau\approx 14925$. In this range of $\epsilon$ the largest $\tau_d\approx 1404$, satisfying the condition $\tau\gg \tau_d$.  Panel b: $A_{\rm max}$ as a function of $\tau$ for $\epsilon\approx 0.0313$.
	In this range the opposite condition $\tau \ll \tau_d$ is satisfied. }
\label{validation}
\end{figure}
%%%%%%%%%%%%%%%%%%%%%%%%%%%%%%%%%%%%%%%%%%%%%

Finally, we should discus the dependence of the response on the distance $\epsilon$ from the point of instability and on
the magnitude of the damping coefficient. Eqs.~(\ref{ampeps}) and (\ref{amptau}) predict that for a fixed $\tau \gg\tau_d$, the maximal
amplitude of the response should be linear in $1/\sqrt{\epsilon}$, and that for fixed (and small) $\epsilon$ and $\tau \ll \tau_d$, it should be linear 
in $\tau$. Both expectations are validated by measuring the square root of the maximal MSD response as shown in Fig.~\ref{validation}.
This maximal value is denoted $A_{\rm max}$.
In panel (a) of that figure we present the dependence on $\epsilon$ for fixed $\tau$ and in panel (b) the dependence
on $\tau$ for fixed $\epsilon$. The results are in excellent agreement with Eqs.~(\ref{ampeps}) and (\ref{amptau}). Together these 
results also validate Eq.~(\ref{ampfinal}).

We reiterate that the present numerics serve just to demonstrate the validity of the theoretical analysis, which
is much more general than the present example. The giant sensitivity to external noise should exist in a variety
of systems in the class studied above, i.e. systems in which the forces appearing in Newton's equations of motion
are not derivable from a Hamiltonian.

\section{Summary and Concluding Remarks}
\label{summary}

The aim of this paper was to present a generic mechanism for high sensitivity to small external perturbations that can trigger
a major event that is associated with a close-by instability in an otherwise stable system. We have in mind remote triggering of earthquakes, but our discussion
is more general, pertaining to physical systems in which the forces are not derived from a Hamiltonian. In such systems, there is a generic instability in which pairs
of complex eigenvalues get born, leading to an exponential growth of any deviation from mechanical equilibrium. We explained that this instability differs from
the standard Hopf bifurcation in which only two modes are involved. Here we need four modes to be involved, making this instability ``less generic". On the other hand
the standard Hopf bifurcation is only sensitive to resonant perturbations. The instability discussed here is sensitive to any perturbation independent of its
frequency, which is in the plane containing the two eigenvectors that coalesce at the instability. It is enough to have a component in the direction
perpendicular to the coalescing eigenvectors to trigger the instability. Thus the perturbation can be ``more generic" than the one
required to trigger a Hopf phenomenology. We have demonstrated this high sensitivity with a simple model of frictional disks that exhibit a transition
from static to dynamical friction. The direct connection to geophysical instabilities and earthquakes needs further study, but in light of the
genericity discussed above we hope that this
paper will motivate such studies in the near future.

Finally we should comment on the issue of ``exceptional
points" in quantum models with non-Hermitian Hamiltonians. The bifurcation mechanism described in this paper is directly, albeit somewhat unexpectedly, related to the notion of exceptional points  known as degeneracies in systems governed by non-Hermitian evolution operators \cite{Moiseyev2011, Heiss2012}. The exceptional points are defined as special degeneracies, where two modes of the system have the same frequencies and modal shapes \cite{Moiseyev2011, Heiss2012, Pick2017}. The exceptional points are responsible for a plethora of counter-intuitive phenomena in open optical systems, such as giant spontaneous light emission \cite{Pick2017}, unidirectional reflection and transmission \cite{Lin2011} or topological mode switching \cite{Doppler2016}. In laminate composites the exceptional points are related to anomalous energy transport phenomena, such as negative refraction, beam steering and splitting \cite{Lustig2019}.
It is clear that the asymmetric operator in our system is a classical counterpart of the non-Hermitian evolution operator. It is also clear that the bifurcation involving the coincidence of frequency and modal shape in the system of frictional disks is analogous to the exceptional points.  In all aforementioned applications, the exceptional points are treated as exotic degeneracies that require precise, possibly multi-parametric tuning of the system.  In contrast, our systems are disordered and have a large number of degrees of freedom. In these systems, in the absence of  viscous damping, the mode coalescence is a generic bifurcation. It is characterized by infinite sensitivity in the linear approximation, due to violation of the mode orthogonality. Small damping preserves the giant sensitivity, while the bifurcation remains generic. Thus, in the dynamics of  forced systems of frictional disks states similar to the exceptional points should be considered as generic and not exceptional.

\acknowledgments

This paper has been supported in part by the ISF joint program with Singapore, the US-Israel
BSF and the Minerva Foundation, Munich, Germany, through the Minerva Center for Aging at the Weizmann Institute. 

\bibliography{Sensitivity}
\end{document}